# Simultaneous control of spectral and directional emissivity with gradient epsilon-near-zero InAs photonic structures

Jae Seung Hwang[1,+], Jin Xu[1,+] and Aaswath P. Raman[1,2*]

[1]Department of Materials Science and Engineering, University of California, Los Angeles, Los Angeles, CA 90095 USA

[2]California NanoSystems Institute, University of California, Los Angeles, Los Angeles, CA 90095 USA

*Corresponding Author: aaswath@ucla.edu

**Abstract**

Controlling both the spectral bandwidth and directional range of emitted thermal radiation is a fundamental challenge in modern photonics and materials research. Recent work has shown that materials with a spatial gradient in their epsilon near zero response can support broad spectrum directionality in their emissivity, enabling high radiance to specific angles of incidence. However, this capability has been limited spectrally and directionally by the availability of materials supporting phonon-polariton resonances over long-wave infrared wavelengths. Here, we design and experimentally demonstrate an approach using doped III-V semiconductors that can simultaneously tailor spectral peak, bandwidth and directionality of infrared emissivity. We epitaxially grow and characterize InAs-based gradient ENZ photonic structures that exhibit broadband directional emission with varying spectral bandwidths and peak directions as a function of their doping concentration profile and thickness. Due to its easy-to-fabricate geometry we believe this approach provides a versatile photonic platform to dynamically control broadband spectral and directional emissivity for a range of emerging applications.

**Introduction**

The ubiquity of thermally-generated light makes its control of fundamental importance in a broad range of applications, from thermophotovoltaics (TPV) (1-3) to thermal imaging (4-6) and infrared sensing (7-9). Far-field thermal emission is typically incoherent both spatially and temporally, resulting in its broad spectrum and omnidirectional nature (10). However, thermal emission into unwanted frequencies and directions greatly limits the efficiency of existing technologies (11). The ability to control the spectral and angular distributions of thermal emission has thus stimulated an active field of contemporary photonics and materials research.

A range of photonic strategies have been investigated to control the spectral and directional nature of thermally generated light emission, including surface plasmon polaritons (12-17), phonon polaritonic structures (18, 19), metasurfaces (1, 11), hyperbolic metamaterials (20, 21) and photonic crystals (22). While significant progress has been achieved in spectrally selective thermal emitters, directional control has proved more challenging. When directional thermal emission has been deliberately engineered, the spectral bandwidth of directional emission is typically narrowband and the emission angle itself changes significantly as a function of wavelength. By contrast broadband directional thermal emission control, where emissivity is highly directional to the same set of angles across a broad bandwidth, has proved to be more challenging to achieve. Recently, strategies to achieve such a capability were proposed theoretically (23, 24), and an alternate approach using polaritonic oxide based 'gradient' epsilon near zero (ENZ) materials was experimentally demonstrated (25).

In gradient ENZ materials, the ENZ frequency varies spatially along a key dimension, typically the thickness of a film. Subwavelength ENZ films are known to support a leaky electromagnetic mode (the Berreman mode) near their longitudinal optical phonon frequencies in the p polarization, whose angular response is controlled by film thickness (26-31). In our recent work, gradient ENZ films were shown to support a broadband Berreman mode that in turn enabled directional thermal emission to the same angular range over 4-micron bandwidths in the long-wave infrared (LWIR) wavelength range (25). However, a key limitation of this approach is that the gradient ENZ response was achieved using complementary phonon-polariton resonances supported by a range of oxides. This in turn meant that the spectral range, bandwidth, and directional characteristic of such thermal emitters is limited by suitable materials available in nature. Furthermore, as the resonance frequencies of the constituent layers are fixed by their intrinsic electron or phonon interactions, there is no avenue for active tuning of spectral or directional emissivity in such films.

Alternatively, the desired ENZ response can be enabled by coupling to free carriers, electrons and holes, in semiconductors. Doped semiconductors can exhibit a Drude metal-like response with their plasma frequency typically at infrared wavelengths and determined by their free carrier concentration (32-40). Due to this capability, prior studies have used doped semiconductor-based platforms to demonstrate Tamm plasmon polariton emitters (41, 42) and dynamic thermal emission control (43-45). Broadband omnidirectional emissivity was also demonstrated using resonances supported in semiconductor nanostructures (13, 14, 46). Doped semiconductors have also been used to demonstrate Berreman mode-driven behavior, including directional emission, but only over a narrow bandwidth of operation (33, 36, 37). Recent work has demonstrated that multiple layers of CdO doped at different doping concentrations can yield multi-spectral narrowband or broadband emitters at the Brewster angle of the substrate (47). However, no prior work has demonstrated the ability to simultaneously tune the spectral bandwidth, spectral range and angular range of a directional emitter. Such a capability would allow one to constrain directional emission to particular angular ranges for arbitrary spectral ranges, a challenging prospect not previously demonstrated.

Here, we present a new platform for tailoring spectral and directional thermal emission with unprecedented control: semiconductor gradient ENZ photonic structures based on epitaxially grown graded doped InAs. We show that the spectral peak and bandwidth of high emissivity can be tuned to a remarkable extent by controlling both the thickness and the doping concentration profile along the depth dimension of the structure. We epitaxially grow and characterize photonic structures demonstrating, by design, high emissivity peaks between 17.5 to 19.5 $\mu$m and between 12.5 to 15$\mu$m respectively. Furthermore, we demonstrate directional tuning of the thermal beaming effect by controlling the total thickness of the structure, where the directionality is constant over the entire high emission bandwidth. We demonstrate directional tuning over 2-micron bandwidths from a 74° peak to a 66° peak. Our III-V semiconductor-based gradient ENZ platform opens up exciting possibilities for novel device concepts (32), such as dynamically controlling the broadband operational wavelength range of directional thermal emission (45), beam steering, as well as broadband non-reciprocal thermal emission control (48).

**Results**

To enable the full potential of gradient ENZ films and demonstrate tailored control of spectral and directional emissivity, we exploited the free carrier response of judiciously designed Silicon-doped InAs thin films where the doping concentration varies spatially in a graded profile along the depth dimension. The first structure we designed and grew by molecular beam epitaxy was composed of InAs thin films of varying thickness with the doping concentration ranging from $1.0\times10^{18}$ cm$^{-3}$ to $1.9\times10^{18}$ cm$^{-3}$ (Fig. 1A). In the first structure, the thickness of the individual layers was 30 nm, and they were epitaxially grown atop a heavily doped InAs layer with a doping concentration of $5.0\times10^{19}$ cm$^{-3}$ and thickness of 300nm (Fig. 1A). The heavily doped InAs layer serves as a reflective mirror supporting the Berreman mode in the ENZ wavelength ranges of the gradient InAs thin films, as it has negative permittivity over the key wavelengths of operation of the gradient ENZ layers. The gradient ENZ InAs structure and the heavily doped InAs layer were both grown on a 50mm (2-inch) semi-insulating GaAs substrate, with a total InAs film thickness of 600nm (Fig. 1B).

To motivate the doping profile chosen, we numerically calculated the real and imaginary parts of the dielectric constant of the constituent semiconductor thin films of the gradient InAs layer for structure 1 (Fig. 1C and Fig. S1). With increasing doping concentration from $1.0\times10^{18}$ cm$^{-3}$ to $1.9\times10^{18}$ cm$^{-3}$, the ENZ wavelength gradually varies from 19.2$\mu$m to 18.2$\mu$m (Fig. 1C) while the permittivity of the heavily doped (n$^{++}$ InAs) layer remains highly negative in this wavelength range, allowing the field to be locally confined in the gradient ENZ thin film. The imaginary part of the permittivity remains small for the thin films at these wavelength ranges enabling maximal directional contrast (Fig. S1). Due to their complementary resonance frequencies that vary gradually along the depth dimension, layering these semiconductor thin films will result in a III-V-based gradient ENZ thin film where its broadband spectral emission is more continuous, and tunable, compared to the previously demonstrated gradient ENZ thin films based on polaritonic oxides (49).

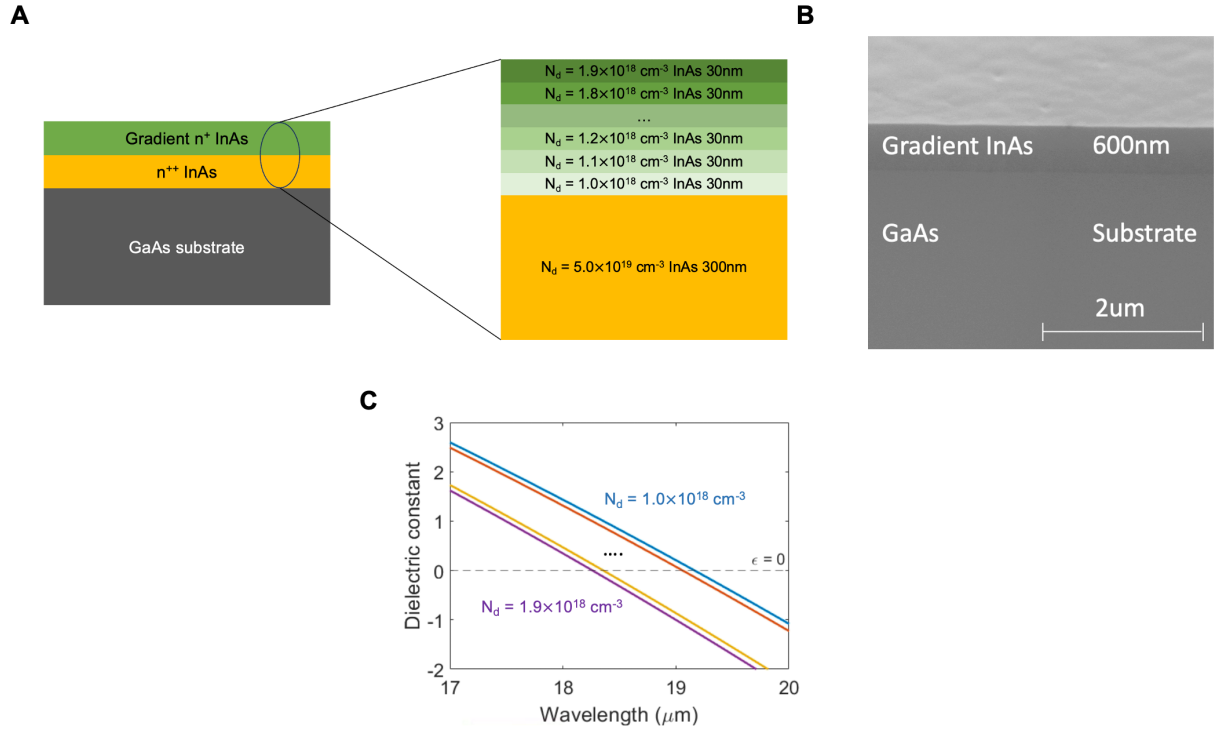

**Figure 1: Configuration of a InAs-based gradient ENZ thin film for broadband directional thermal emission.** (**A**) Schematic of thermal emitter enabled by a semiconductor gradient ENZ thin film. The doping concentration of the constituent semiconductor thin films varies spatially along the depth dimension with a range from $1.9\times10^{18}$ cm$^{-3}$ to $1.0\times10^{18}$ cm$^{-3}$. The thickness of the individual layers is 30 nm. (**B**) SEM image of the experimentally fabricated multi-layer InAs film structure, with labels identifying the materials used and layer thicknesses. (**C**) Numerically calculated real part of the permittivity (dielectric constant) of the gradient doped semiconductor thin films with a doping concentration range from $1.9\times10^{18}$ cm$^{-3}$ to $1.0\times10^{18}$ cm$^{-3}$. The ENZ wavelength varies with different doping concentration in a continuous range.

We show in Fig. 2A the emissivity spectra of the fabricated structure varying with angle and wavelength in the p-polarization, as measured by a FTIR spectrometer. The spectrum exhibits a strong high emissivity resonance band spanning continuously along the wavelength range from 18.2$\mu$m to 19.2$\mu$m associated with the ENZ wavelength range of the constituent semiconductor thin films of the gradient ENZ film. The emissivity spectra also show a strongly angularly selective behavior in which the high emission angular range where the overall p-polarized emissivity is above 0.4 throughout the 17.5 to 19.5 $\mu$m range is centered at 76° (Fig. 2A).

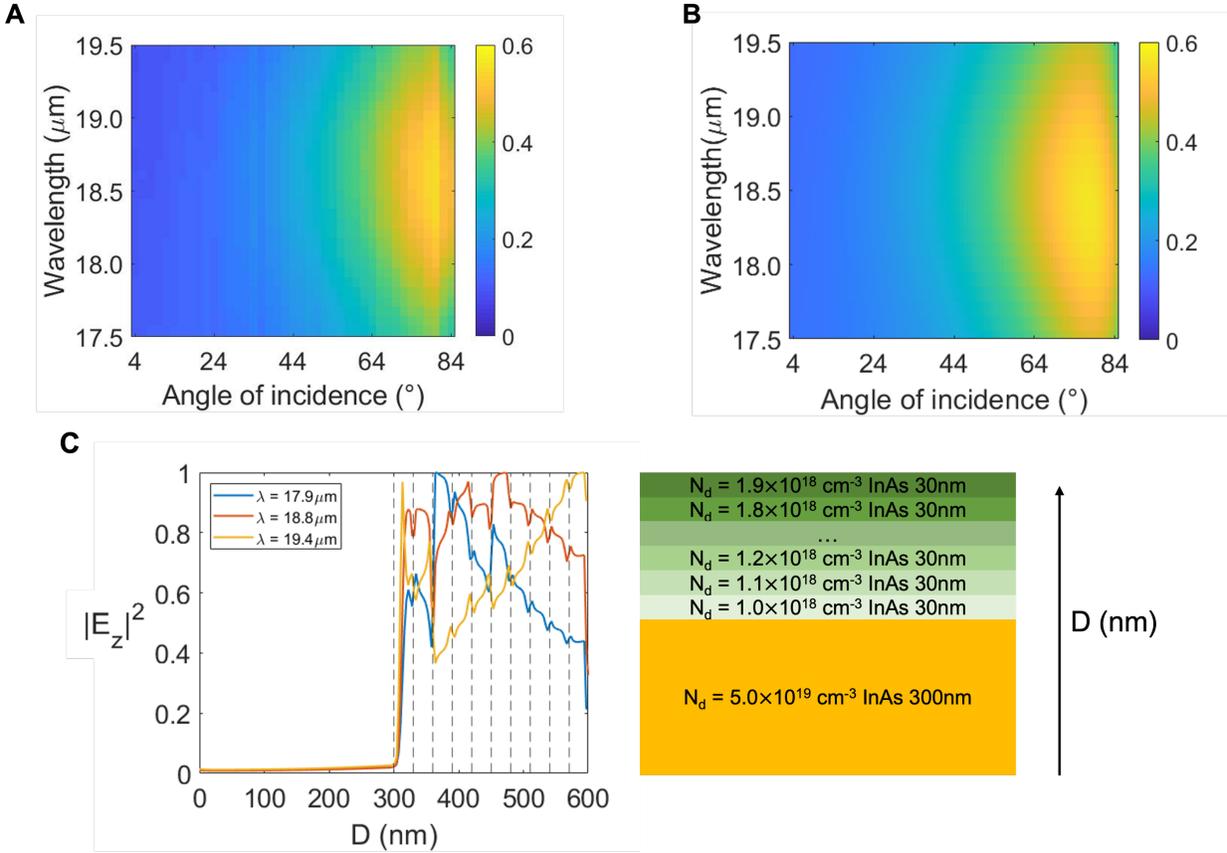

**Figure 2: Comparing experimental measurements with simulated results.** (**A**) Experimentally measured emissivity spectra varying with angle and wavelength of the fabricated multi-layer semiconductor film structure in p-polarization. (**B**) Simulation result of the emissivity spectra of the structure for p-polarization using the transfer matrix method. The simulation result shows a broadband angular selective behavior in emissivity and is in good agreement with the experimental measurement. (**C**) Simulated electric field intensity distribution for the peak emissivity angle illumination at wavelengths ($\lambda$) of 17.9 $\mu$m, 18.8 $\mu$m and 19.4 $\mu$m. The layers with higher doping concentrations support the Berreman mode peak at lower wavelengths, whereas the thin films with relatively lower doping concentrations support the optical mode at higher wavelengths corresponding to their ENZ wavelengths.

To further validate our experimental results, we next show the simulated emissivity spectra calculated using the transfer matrix method (Fig. 2B). The simulation results agree well with the experimental results, with both exhibiting strong directional emissivity throughout a broad bandwidth between 17.5 to 19.5$\mu$m. The operation range as well as the angles of operation and peak angle of incidence agree well with the experimental data. To elucidate the origin of this behavior, the electric field intensity distribution for the peak emissivity angle illumination at wavelengths ($\lambda$) of 17.9$\mu$m, 18.8$\mu$m and 19.4$\mu$m was studied using electromagnetic simulations (Fig. 2C). The constituent layers of the gradient ENZ thin film with higher doping concentrations support the Berreman mode at smaller wavelengths ($\lambda$=17.9$\mu$m) of the operation range and the field intensity profile shows larger field enhancement localized at the upper layers at these wavelengths. The layers with relatively lower doping concentrations support the optical mode at larger wavelengths ($\lambda$=19.4$\mu$m) and the field intensity profile shows a larger field enhancement localized at the lower layers at these wavelengths. At wavelengths near the peak emission ($\lambda$=18.8$\mu$m), the field is enhanced throughout the bulk of the gradient ENZ thin film. This localized field enhancement can be easily understood by considering two adjacent medium 1 and 2 characterized by a permittivity of $\varepsilon_1$, $\varepsilon_2$ and electric field of $E_1$, $E_2$, respectively. The z-component of electric displacement field $\mathbf{D} = \varepsilon\mathbf{E}$ is continuous at the interface of the two adjacent mediums so that $E_{z,2} = (\varepsilon_1/\varepsilon_2) E_{z,1}$ (26). It follows that the field $E_{z,2}$ is enhanced if $\varepsilon_2$ approaches zero near the ENZ wavelength of medium 2 (Fig. 1C). The heavily doped InAs layer remains highly reflective at these wavelength ranges exhibiting close to zero electric field intensity.

We next show that the fabricated based gradient ENZ materials enable tunability of the spectral range in absorption and emission by controlling the doping concentration range of the gradient ENZ thin film. We fabricated a second semiconductor gradient ENZ structure (structure 2) with the doping concentration of the gradient ENZ thin film ranging from 2.0×10$^{18}$ cm$^{-3}$ to 4.5×10$^{18}$ cm$^{-3}$. The thickness of the individual layers was 50 nm so that the overall total thickness of the gradient ENZ thin film was 300 nm, identical to structure 1. We measured the emissivity spectra varying with angle and wavelength of the fabricated multi-layer semiconductor film structure in p-polarization (Fig. 4A). For structure 2, the spectrum exhibits high emissivity in the wavelength range from 12.5$\mu$m to 15$\mu$m corresponding to the ENZ wavelength range of the constituent semiconductor thin films of the gradient ENZ layer (Fig. S1). The emissivity spectra also exhibit a strongly directional behavior in which the high emission angular range where the overall p-polarized emissivity is above 0.4 throughout the 12.5 to 15$\mu$m range is centered at 74° (Fig. 4A), similar to structure 1, but over a different spectral range.

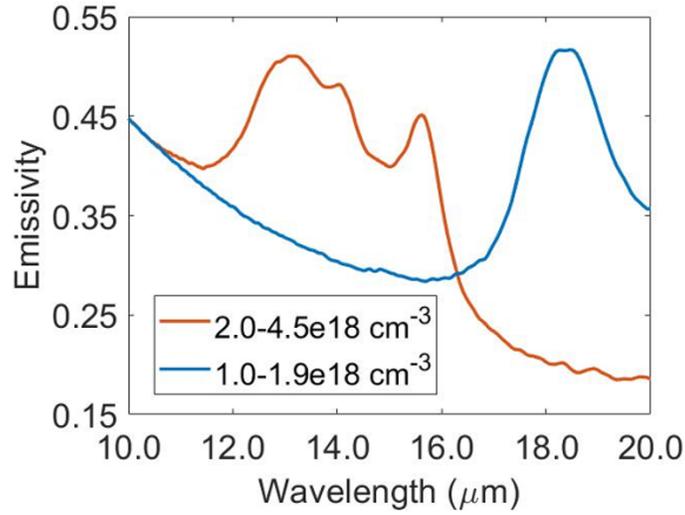

**Figure 3: Tuning the spectral peak and bandwidth of the broadband directional thermal emitter by controlling the doping concentration profile of the gradient ENZ layer.** Comparison of the measured emissivity in p-polarization between the two gradient ENZ structures with different doping concentration ranges at a 76° angle of incidence. The working range of structure 2 is situated at lower wavelengths and the bandwidth is wider compared to structure 1.

We simulated the emissivity spectra using the transfer matrix method and the simulation result of the fabricated structure agrees well with the experimental result, exhibiting angular regions of high and low emissivity throughout a broad bandwidth between 12.5 to 15$\mu$m. The operational range as well as the angles of operation and peak angle of incidence agree well with the experimental data (Fig. 4B). To better characterize the tunability of the operation range, we plotted the spectral emissivity at peak emissivity angles for both structure 1 and 2 (Fig. 3). We clearly see that the two structures exhibit different operation ranges corresponding to the ENZ wavelength range of the constituent layers of the gradient ENZ thin film. We also observe that the operational bandwidth of structure 2 is wider than structure 1 due to its larger spatial gradient of the doping concentration of the gradient ENZ thin film. This highlights the remarkable control over spectral emissivity bandwidth that III-V-based gradient ENZ structures can provide by controlling the doping concentration profile of the gradient ENZ layer.

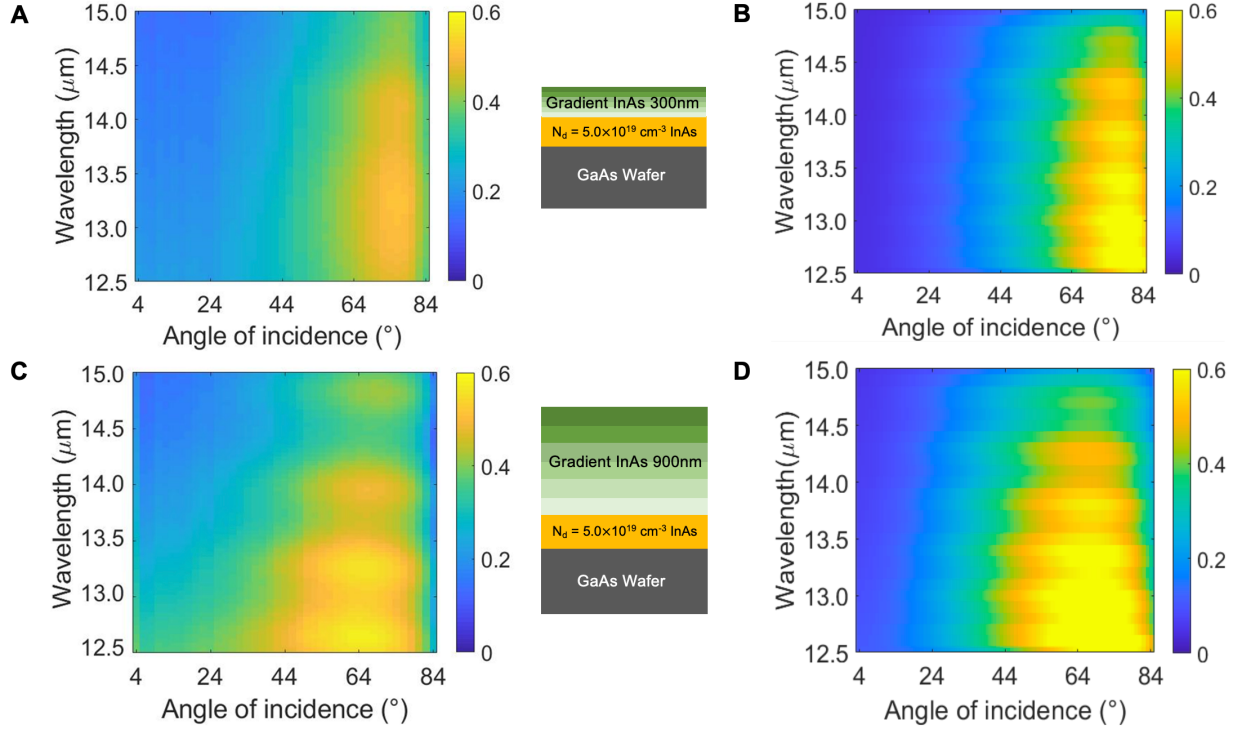

**Figure 4: Tuning the directionality of the broadband directional thermal emitter by controlling the total thickness of the semiconductor gradient ENZ thin film.** (**A**) and (**C**) Measured emissivity spectra varying with angle and wavelength of the two fabricated multi-layer semiconductor film structures in p-polarization with the same doping concentration range from $2.0\times10^{18}$ cm$^{-3}$ to $4.5\times10^{18}$ cm$^{-3}$, but different total thickness of 300nm and 900nm. (**B**) and (**D**) Simulation results of the emissivity spectra of the structure for p-polarization using the transfer matrix method. The simulation results are in good agreement with the experimental measurements.

We next demonstrate that, in addition to tuning the spectral bandwidth, one can simultaneously tune the directional response of semiconductor thin film based gradient ENZ materials by controlling the total thickness of the gradient ENZ layer. We fabricate an additional gradient ENZ structure (structure 3) with the doping concentration range of the gradient ENZ thin film ranging from $2.0\times10^{18}$ cm$^{-3}$ to $4.5\times10^{18}$ cm$^{-3}$, identical to structure 2. Here, the thickness of the individual layers was increased to 150 nm so that the total thickness of the gradient ENZ thin film was 900nm. For structure 3, the emissivity spectrum exhibits high emissivity in the same wavelength range as structure 2 (from 12.5$\mu$m to 15$\mu$m) (Fig. 4C), corresponding to the ENZ wavelengths of the constituent semiconductor thin films of the gradient ENZ layers. The emissivity spectra show a strong, broadband directional response. however, the directional peak, where the overall p-polarized emissivity is above 0.4 throughout the 12.5 to 15 μm range, was centered at 66° (Fig. 4C). The simulation results using the transfer matrix method also shows a similar shift in the peak emission angle towards normal incidence as compared to structure 2 (Fig. 4B and D). For structure 2, we measured an average emissivity of > 0.4 in the p-polarization over the wavelength range of operation (12.5 to 15 μm) between 66° to 80° (Fig. 5A, red line). Outside of this angular range, the average emissivity drops below 0.3 at 50°. By contrast, for structure 3, over the same wavelength range of operation (12.5 to 15 μm) for p-polarization, we measure an average emissivity > 0.4 between 58° and 74° (Fig. 5A, black line). Outside of this angular range, the average emissivity drops to below 0.3 at 40°. By controlling the doping concentration gradient and thickness we are thus able to achieve exquisite and tailored control over directional range of high emissivity while maintaining that directional response over a broad bandwidth by design. As is seen in Fig. 5B, this response originates fundamentally from the shift in the broadband Berreman mode supported by the thicker gradient ENZ InAs photonic structure relative to the thinner one.

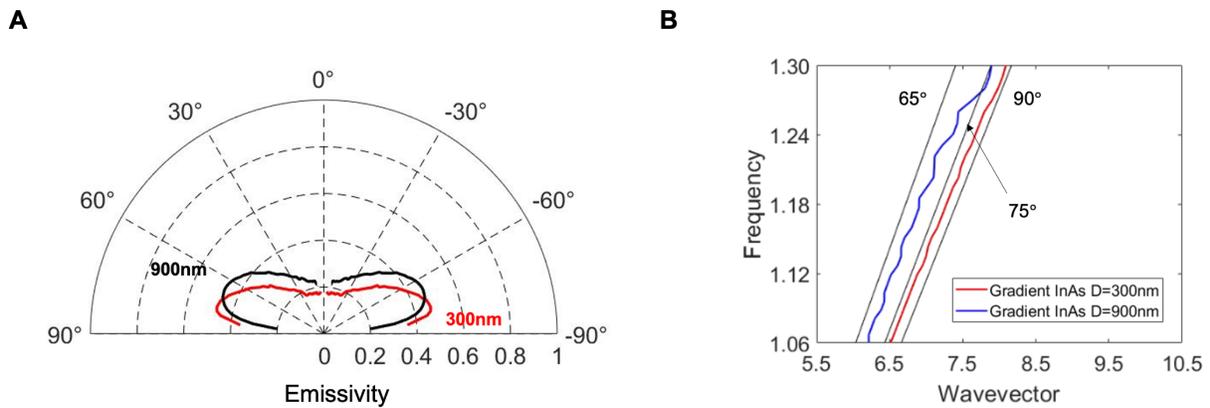

**Figure 5: Directional response of the two gradient ENZ thin films with different total thickness.** (**A**) Polar plot of the measured average emissivity over a broad wavelength range of operation (12.5 to 15 μm) varying with angle of incidence for p-polarization in structure 2 and 3, with peak emissivity observed at 74° and 64°, respectively. Larger total thickness of the gradient ENZ thin film shifts the high emissivity angle towards normal incidence. (**B**) The dispersion relation of structure 2 and 3 throughout the frequency range of interest. The broadband Berreman mode moves towards normal incidence as the total thickness of the gradient ENZ layer increases from 300nm to 900nm, which agrees with (**A**).

**Conclusions**

We have demonstrated a III-V semiconductor-based platform that offers unprecedented control of spectral and directional emissivity over infrared wavelengths. We show that our gradient ENZ scheme has the capability to tune the spectral bandwidth and spectral range of a directional emitter arbitrarily and simultaneously. Furthermore, by controlling the total thickness of the gradient ENZ structure, we show that the angular response can be simultaneously tuned as well. We emphasize that our directional emitters have emissivity that is highly directional to the *same* set of angles across an arbitrary bandwidth. Such a capability allows one to constrain directional emission to particular angular ranges for arbitrary spectral ranges, which is a challenging, but enabling capability in a broad range of applications such as thermophotovoltaics (50), radiative cooling (51) and waste heat recovery (52). It is noteworthy that the lithography/patterning free, chip-scale geometry we have used suggests that these semiconductor gradient ENZ structures can be integrated with other photonic structures, opening up intriguing possibilities for controlling absorption and emission. Ultimately, since the demonstrated semiconductor gradient ENZ photonic structures are driven by free carrier concentrations which can be electrically tuned (32), we believe that this configuration provides a suitable platform for dynamic control of broadband directional thermal emission.

**Acknowledgements**

This material is based upon work supported by the National Science Foundation under grant no. ECCS-2146577 and the Sloan Research Fellowship (Alfred P. Sloan Foundation).